\documentclass[aps,prc,floatfix,preprint,longbibliography,nofootinbib,showkeys]{revtex4-1}

\usepackage{amsmath,amssymb,amsfonts,graphicx,pifont,dcolumn}
\usepackage{epstopdf}
\usepackage{color}
\usepackage{cancel}
\usepackage{ulem}
\usepackage{siunitx}
\usepackage{bm}
\usepackage[colorlinks,linkcolor=blue,citecolor=blue]{hyperref}   

\RequirePackage{slashed}

\def\xP{x_{\!I\!\!P}}

\def\bmDelta{\mbox{\boldmath$\Delta$}}\def\bb{{\bf b}}
\def\br{{\bf r}}
\def\cc{{\{c\}}}

\def\r{{\boldsymbol r}}
\def\x{{\boldsymbol x}}
\def\y{{\boldsymbol y}}

\def\b{{\boldsymbol b}}

\newcommand{\rmd}{{ \rm d}}
\newcommand{\rme}{{\rm e}}

\def\bra#1{\langle#1\vert}
\def\ket#1{\vert#1\rangle}

\newcommand{\bit}{\begin{itemize}}
\newcommand{\eit}{\end{itemize}}

\def\be{\begin{equation}}
\def\ee{\end{equation}}
\def\bea{\begin{eqnarray}}
\def\eea{\end{eqnarray}}

\begin{document}


\title{Effect of dipole size fluctuations on diffractive photo-production of vector mesons}

\author{Jean-Paul~Blaizot}
\affiliation{Institut de Physique Th\'eorique, Universit\'e Paris Saclay, CEA, CNRS, F-91191 Gif-sur-Yvette, France}\email[]{jean-paul.blaizot@ipht.fr}

\author{Marco~Claudio~Traini} 
\affiliation{INFN - TIFPA, Via Sommarive 14, I-38123 Trento-Povo, Italy}
\affiliation{Dipartimento di Fisica, Universit\`a degli Studi di Trento, 
Via Sommarive 14, I-38123 Trento-Povo, Italy}

\begin{abstract}We consider the diffractive photo-production of vector mesons on a proton, in the dipole model. We take into account the effect of the fluctuations of the the dipole size, whose magnitude is controlled by the overlap between the photon and the vector meson wave functions. Our predictions for the incoherent diffractive cross section, obtained  within the Impact Parameter Saturation Model (IPSat), are shown to be in excellent agreement with the HERA data on $J/\Psi$ photo-production, down to very low momentum transfer, where the dipole size fluctuations are shown to play an essential role. This study complements, without introducing any additional parameter, previous treatments of incoherent diffractive processes in terms of  fluctuations of the proton shape, by adding another source of ``geometrical'' fluctuations, namely those coming from the splitting of the photon into color dipoles of various sizes.

\end{abstract}

\maketitle

\section{\label{sec:intro} Introduction}

Deep inelastic scattering of leptons on hadrons is a tool of choice to obtain information on the  structure of hadrons, or nuclei, in terms of partons. At high energy, and in an appropriate frame, it is often convenient to picture the interaction as a two-step process, whereby  the virtual photon exchanged between the lepton and a hadron  first splits into a quark-antiquark dipole, which later interacts strongly with the hadron. When the interaction of the dipole with the hadrons involves partons that carry a very small fraction $x$ of the hadron momentum, such an approach gives access to the gluon density in the hadron and its possible saturation property \cite{Mueller1990}.  This picture is at the basis of the color dipole model \cite{dip1,dip4}, which exists in various incarnations (see e.g. \cite{GBW1,GBW2,IPSat2003,IPSat2002,IPSat2006}).  

Dipole models are also useful to understand diffractive events, which represent a large fraction of the electron-hadron cross section at  high energy (for a recent review see \cite{diffraction_review}). Many studies have been carried out within the last two decades using such models, with in particular  the goal of getting evidence for saturation effects in the diffractive production of different final states~\cite{GBW1,GBW2,IPSat2003}. The vector meson production off nucleons and nuclei~\cite{VMp1} is a particularly  clean process since the final state is unambiguously identified by a large rapidity gap. Such  processes are  strongly related to the underlying QCD dynamics and are sensitive to  the gluon content of the target \cite{QCD_HE}. Among the many existing models, a  popular one is the impact parameter dependent saturation model (IPsat)~\cite{IPSat2002,IPSat2003,IPSat2006,IPSat2013,IPSat2017} whose basic parameters are determined through fits to available high precision data of  deep inelastic scattering of electrons on protons   as measured at HERA~\cite{HERA2010,HERA2015}  (see ref.~\cite{IPSat2018} for a recent review). This is the model that we shall mainly use in this work.

 However, our main interest in this paper is not the physics of saturation, but rather that of fluctuations that dominate the  diffractive cross section at moderate momentum transfer.  In this context, incoherent diffraction  plays a prominent role  because it is sensitive to the transverse fluctuations of the gluon field in the target. To study these fluctuations, we shall rely on the formalism of diffraction eigenstates \cite{GoodWalker1960,MiettinenPumplin1978}. The states of the dipole, defined by the coordinates of the quark and antiquark provide an example of such eigenstates. This is because, when the dipole picture is a valid one,  the lifetime  of the dipole  is long as compared to the size of the hadron, and the dipole coordinates remain frozen during the scattering. The coordinates of the sources of the color field of the target, also frozen during the collision,  provide another example of diffraction eigenstate \cite{MiettinenPumplin1978}. In this paper we follow the phenomenological approach initiated in Ref.~\cite{M&Schenke2016} and assume that a dominant source of fluctuations at moderate momentum transfer is that  due to the random location of the valence quarks in each collision. Each valence quark is carrying its cloud of gluons, and the fluctuation of their locations entails a corresponding fluctuation of the gluon field in the proton. We shall stick to this simple picture, being aware of the existence of other models where this  relation between the gluon field and the valence quarks is relaxed, so-called ``hot spot'' models which can be extended to momentum transfers higher than those we shall consider in this paper (see e.g.  \cite{Demirci:2022wuy,Kumar:2021zbn,Albacete:2016pmp} and for a review \cite{MantysaariFluct:2020}). 

The main novelty of the present work concerns the treatment of the fluctuations of the dipole size. Such fluctuations result, event-by-event, from the quantum mechanical process of the photon splitting. Thus one may attach a probability to the splitting, given in principle by the photon wave function. In the case of vector meson production, a subtlety arises due to the fact that what plays the role of a probability is the overlap of the photon wave function with that of the vector meson \cite{MunierStastoMueller2001}. We shall elaborate on this feature and propose a  treatment of geometrical fluctuations that include those of the dipole size. We shall see that such fluctuations account correctly for the incoherent cross section in the region of  small momentum transfer where the fluctuations of the proton shape die out. These dipole size fluctuations, which translate into  fluctuations of the dipole cross section, bear some analogy with the fluctuating cross sections introduced in Refs.~\cite{Blaettel:1993rd, Blaettel:1993ah}. They could also be interpreted as fluctuations of the saturation momentum \cite{M&Schenke2016}. However the simple treatment of dipole size fluctuations suggested in the present paper appears more natural and does not require any new ingredient beyond those already fixed in the model. Note finally that we shall ignore other types of fluctuations, namely those corresponding to gluon emissions \cite{marquet_unified_2007}, intrinsic fluctuations of the color field within the hot spots \cite{SCHLICHTING2014313}, as well as the effects of small-$x$ evolution  \cite{Liou_2017}.

 The outline of the paper is as follows. In the next section we briefly review the formalism commonly used to calculate the vector meson production in the dipole model. Elaborating on the formalism of diffraction  eigenstates \cite{GoodWalker1960,MiettinenPumplin1978}, we recall in Sect.~\ref{flucttarget} how fluctuations of the locations of the sources of the gluon field contribute to the incoherent diffractive cross section. We give a simple argument, based on the relation between these geometrical fluctuations and the density-density correlation function,  why the contribution of these fluctuations to the incoherent cross section vanishes at small momentum transfer. In Sect.~\ref{sec:projfluct}  we develop our suggestion for estimating the magnitude of the fluctuations of the dipole size. The following section, Sec.~\ref{sec:EbyEA}, provides details on the numerical calculations and discusses our results and their comparison with HERA data. The last section summarizes our conclusions. Various appendices collect technical details on the calculation.

\section{Vector meson production in the dipole model}

The amplitude ${\cal A}_{T,L}^{\gamma^* p \to V p'}({\xP},Q^2, \bmDelta)$ for diffractive vector meson production on a proton assumes the form \cite{IPSat2006}
\bea\label{amplitudeA}
&&{\cal A}_{T,L}^{\gamma^* p \to V p'}({\xP},Q^2, \bmDelta)   \nonumber \\
&=& i\int d^2 {\bf r}   \int_0^1  \frac{\rmd z }{4 \pi} \,  \Psi_V^* (z,r,Q^2){\cal A}_{q\bar q}({\xP},Q^2, \r,\bmDelta)\Psi_{T,L} (r,z,Q^2).
\label{eq:A_exclusive0}
\eea
The subscripts $T$ and $L$ refer to the transverse and longitudinal polarizations of the exchanged virtual photon $\gamma^*$,
${\xP}=(P-P')\cdot q/(P\cdot q)$ is the fraction of the longitudinal momentum of the proton involved in the exchange,  and the total momentum transfer (squared) is $t = -(P'-P)^2$ with $P$ and $P'$ the initial and final proton four-momenta. The photon-proton scattering is characterized by a total center-of-mass-energy squared $W^2 = (P+q)^2$, ($Q^2 = -q^2$). Finally ${\bmDelta}=(P'-P)_\perp$ is the transverse momentum transfer. The normalisation of the amplitude is such that the coherent cross section is given by \cite{MunierStastoMueller2001}
\bea
 {d \sigma_{T,L}^{\gamma^* p \to V p'} \over dt} = {(1+\beta^2) \over 16 \pi}
 \left\vert  {\cal A}_{T,L}^{\gamma^* p \to V p}({\xP},Q^2, \bmDelta)\right\vert ^2.
\label{eq:dxsection_coh1}
\eea
The factor $(1+\beta^2)$ in Eq.~(\ref{eq:dxsection_coh1})  is discussed in Appendix~\ref{sec:app2} together with other phenomenological corrections. 
\begin{figure}[h!]
\centering\includegraphics[width=0.75\columnwidth,clip=true,angle=0]{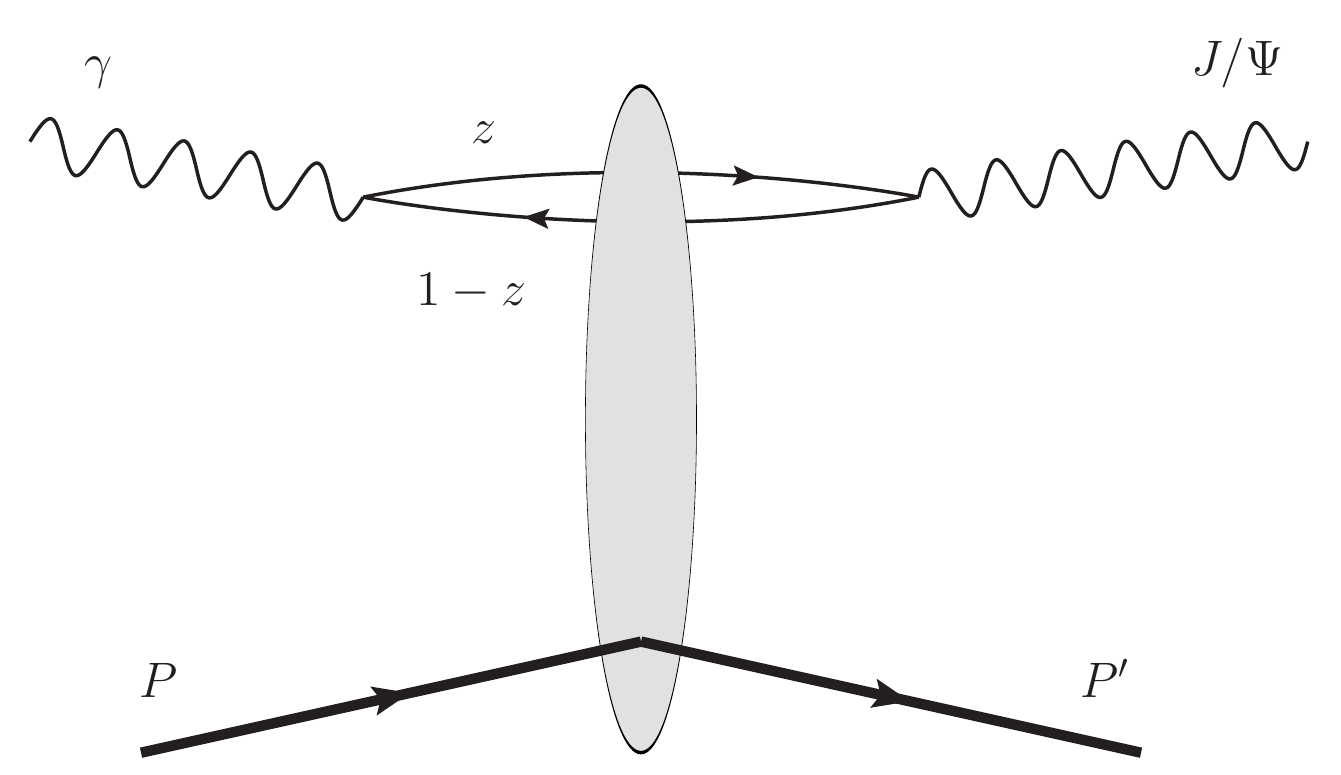}
\caption{  Photo-production of a $J/\Psi$ in the dipole model}
\label{fig:gammap}
\end{figure}

As indicated in Eq.~(\ref{amplitudeA}), see also Fig.~\ref{fig:gammap}, the $\gamma^*\,p$ scattering proceeds through three steps: (1) First, the incoming virtual photon fluctuates into a pair of nearly on-shell quark and antiquark; the transverse size of the pair is $\br$ and the quark carries a fraction $z$  of the photon's light-cone momentum. At high energy the lifetimes of such fluctuations are much longer than the interaction time, so that the transverse size and orientation of the color dipole can be considered frozen during the scattering process  \cite{dip1,dip2}.  The splitting of the photon is described by the virtual photon wave function $\Psi_{T,L} (r,z,Q^2)$, which can be calculated  perturbatively in QED (see e.g. Refs.~\cite{NNPZ94_97_1,NNPZ94_97_2,QCD_HE}). 
 (2) The $q \bar q$ pair scatters elastically on the proton, with an amplitude $i{\cal A}_{q \bar q}$. The factor $i$ here takes into account the fact that the amplitude is essentially imaginary (that is, with this factor, ${\cal A}_{q\bar q}$ is real). (3) Finally the scattered dipole recombines to form a final state, in the present case the vector meson with wave function $\Psi_V(r,z,Q^2)$.

For simplicity, in this paper we shall focus on photo-production, $Q^2=0$ and only the transverse part of the photon wave function contributes. We  then simplify the notation and write the amplitude as 
\bea
{\cal A}^{\gamma p \to V p'}({\xP},\bmDelta) = i  \int d^2 {\bf r}   \int_0^1  {\rmd z \over 4 \pi} \, [ \Psi_V^* \Psi_\gamma ]_{(r,z)}\,{\cal A}_{q\bar q}({\xP},\r, \bmDelta).
\label{eq:A_exclusive1}
\eea
 The overlap factor $[ \Psi_V^* \Psi_\gamma ]_{(r,z)}$ of the wave functions of the photon and the vector meson is discussed in Appendix~\ref{sec:overlapPSI}. In the next subsection, we recall briefly the calculation of the amplitude ${\cal A}_{q\bar q}({\xP},\r, \bmDelta)$ in the dipole model.

\subsection{\label{sec:diff_states}{The dipole cross section}}

In this and the next subsection, we consider the scattering of a color dipole of fixed size on a complex target, such as a proton or a nucleus. We start with the scattering on a  proton, and move to the impact parameter representation, defining (temporarily, see Eq.~(\ref{eq:sqqintegral}) below)
\be\label{calADelta}
{\cal A}_{q\bar q}(\xP,\r,\bmDelta)= \int \rmd^2 \b\, \rme^{-i\b\cdot \bmDelta} {\cal A}_{q\bar q}(\xP,\r,\b).
\ee

At high energy, the scattering amplitude of the dipole is purely imaginary (it is in the explicit two gluon exchange approximation to be used shortly), so that the ${\cal S}$-matrix is real and ${\cal S}_{q\bar q}=1-{\cal A}_{q\bar q}$. In the eikonal approximation the dipole ${\cal S}$-matrix is given by the average of the product of two Wilson lines that represent the propagation of the quark and the antiquark in the color field of the proton:
\be\label{Sbr}
{\cal S}_{q\bar q}(\xP,\b,\r)= \frac{1}{N_c} \langle {\rm Tr}\left[ U^\dagger(\b+\r/2) U(\b-\r/2)\right]\rangle,
\ee
the average represented by the angular brackets being made over the color field of the proton  (see e.g. \cite{theo4}). This average provides the $x$-dependence of ${\cal S}_{q\bar q}(x,\b,\r)$.
A simple calculation, in the two-gluon exchange approximation, allows us to relate ${\cal S}_{q\bar q}(\xP,\b,\r)$ to the gluon density $\xP g(\xP)$, 
\be\label{Swfl} 
{\cal S}_{q\bar q}(\xP,\b,\r)=1-\frac{1}{2} \sigma_{\rm dip}(\xP,r) T_G(b),
\ee
where $\sigma_{\rm dip}(\xP,r)$,
the total dipole-proton cross section, is given by \cite{Blaettel:1993rd}
\be\label{sigmadip}
\sigma_{\rm dip}(\xP,r)=\frac{\pi^2}{3} r^2 \alpha_S(\mu^2) \, \xP g(\xP,\mu^2).
\ee
The scale $\mu^2$ in the gluon density will be specified later (see Eq.~(\ref{eq:mu}) in the next section). $T_G(\b)$ is the average transverse density profile of the proton, normalized such that $\int \rmd^2 \b \,T_G(\b)=1$. For a spherical proton, which has been assumed in writing Eq.~(\ref{Swfl}),  the average profile depends only on $b=\vert \b\vert$. Simlarly $\sigma_{\rm dip}$ in Eq.~(\ref{sigmadip}) is averaged over the orientation of the dipole and depends only on $r=\vert\r\vert$. In writing Eqs.~(\ref{Sbr}) and (\ref{Swfl}), we have assumed that the total gluon density is smoothly distributed in the transverse plane according to the function $T_G(b)$. A more elaborate treatment will be presented at the end of the next subsection. At this point, we note that the total cross section is equal to the imaginary part of the  forward $q\bar q$ scattering amplitude, that is (omitting to indicate the $\xP$ dependence to simplify the notation)
\be
\sigma_{\rm dip}(r)={\cal A}_{q\bar q}(\r,\bmDelta=0)=\int \rmd^2\b \,{\cal A}_{q\bar q}(\r,\b)=2\int \rmd^2 \b\left[1-{\mathcal S}_{q\bar q}(\r, \b)  \right], 
\ee
where the last equality involves the standard eikonal expression of the total cross section in terms of the ${\cal S}$-matrix. This suggests the identification \cite{MunierStastoMueller2001}
\be\label{dsigmad2b}
{\cal A}_{q\bar q}(\b)=2\left[1-{\mathcal S}_{q\bar q}(\r, \b)  \right]\equiv\frac{\rmd \sigma_{q\bar q}}{\rmd^2\b}.
\ee

The two-gluon exchange calculation is valid in the weak field limit, or equivalently when the gluon density is small enough (also called dilute regime). In the dense regime, multiple scattering need to be taken into account. There are several, more or less equivalent, ways to do that (see e.g. \cite{theo4}), the net result being simply the exponentiation of the ${\cal S}$-matrix, that is 
\be
{\cal S}_{q\bar q}(\r,\b)=\exp\left(  -\frac{1}{2}\sigma_{\rm dip}(r) T_G(b)\right).
\ee
In this case, the $q\bar q$ cross section at a given impact parameter $\b$, Eq.~(\ref{dsigmad2b}), takes the form
\be\label{sigmadipexp}
\frac{\rmd \sigma_{q\bar q}}{\rmd^2\b}=2\left[ 1-\rme^{ -\frac{1}{2}\sigma_{\rm dip}(r) T_G(b)}\right].
\ee
This quantity plays the role of the eigenvalue of the scattering matrix, as will be made more visible shortly. The scattering eigenstates are the states localized at the transverse positions of the quark and the antiquark, which positions are frozen during the scattering.  All the information about the proton is contained in the gluon density $\xP g(\xP)$ and the average transverse density profile, $T_G(b)$. The dependence on the dipole size is explicit in the dipole cross section, $\sigma_{\rm dip}\propto r^2$. It is then convenient to write  ${\cal S}_{q\bar q}$ in Eq.~(\ref{sigmadipexp}) as
\be\label{SQs}
 {\cal S}_{q\bar q}(r,b)=\exp\left(  -\frac{Q_s^2(b) r^2}{4}\right),
 \ee
where 
 \be
 Q_s^2(b) \, r^2= 2 \sigma_{\rm dip}(r) T_G(b).
 \ee
 Here $Q_s(b)$ is the saturation momentum, defined as a function of the impact parameter. 
  The saturation momentum fixes the scale of what is meant by ``small'' or ``large'' dipole, or equivalently what is, for a fixed dipole size, a dilute or a dense system. Eq.~(\ref{SQs}) exhibits the interplay between  the properties of the projectile (e.g. the size of the dipole) and those of the target (the gluon density that enters the saturation momentum) in the scattering process (see the discussion at the end of Sect.~\ref{sec:rFluct}).

\subsection{Fluctuations in the target}\label{flucttarget}

The average profile function $T_G(b)$ introduced above is an average quantity, related to the local average gluon density in the proton. However such a quantity may fluctuate event-by event. Such fluctuations are responsible for the incoherent contribution to the diffractive cross section. To analyze more clearly this important source of fluctuations, we shall first consider the scattering of a dipole on a nucleus \cite{Caldwell:2010zza}, leaving for the moment the description of the proton untouched.  At the end of this subsection, we shall discuss analogous fluctuations for the proton, considered as a composite object made  of valence quarks. 

Consider a nucleus made of $A$ nucleons. The density profile, that is the density in the transverse plane (after integration of the three dimensional nucleon density over the longitudinal coordinate) is given by the convolution of the nucleon density with the gluon density in the proton, i.e., 
\be
\hat T_A(\b)=\sum_{i=1}^A T_{G}(\b_i-\b)=\int \rmd^2 \x \,\hat\rho(\x) T_{G}(\x-\b),
\ee
where $\b_i$ denotes the location of nucleon $i$ and $\b$ is the impact parameter of the dipole (defined at this point at mid distance between the quark and the antiquark, see Eq.~(\ref{Sbr})). $\hat T_A(\b)$  is a one-body operator in the nucleon variables, and $\hat \rho(\x)$ denotes the nucleon density operator in the transverse plane:
\be
\hat \rho(\x)=\sum_{i=1}^A \delta(\x-\b_i).
\ee
 The matrix element $\bra{\Psi_0}\hat T_A(\b) \ket{\Psi_0}=\langle \hat T_A(\b)\rangle $  is easily evaluated in terms of the average density $\langle \hat \rho(\x)\rangle=\bra{\Psi_0}\hat \rho(\x) \ket{\Psi_0}$, 
$
\langle \hat T_A(\b) \rangle=\int \rmd^2\x \,\langle \hat\rho(\x)\rangle T_{G}(\x-\b).
$ Note that $\int\rmd^2\b \,\hat T_A(\b)=A$. 

The expressions (\ref{Swfl}) and (\ref{dsigmad2b}) giving the cross section at a given impact parameter in the weak field limit  generalize  into 
\be\label{dsigd2bTA}
\frac{\rmd \sigma_{q\bar q}} {\rmd^2\b}=\sigma_{\rm dip}(r)  \hat T_A(\b),
\ee
which is now an operator in the nucleon variables. This operator is diagonal in the basis of states $\ket{\b_1,\cdots,\b_A}$, where the $\b_i$'s are the positions of the individual nucleons, frozen during the collision process: these states can be considered as diffraction eigenstates \cite{MiettinenPumplin1978}. The elastic amplitude is obtained by taking the average of this cross section in the ground state wave function, which amounts to  average over the distribution of the positions $\b_i$. At a given momentum transfer  $\bmDelta$, the  coherent diffractive cross section of a dipole on a nucleus is then proportional to 
\be\label{coherentSigTA}
\int \rmd^2\b \,\rmd^2\b' \rme^{-i\bmDelta\cdot(\b-\b')} \left\langle\frac{\rmd \sigma_{q\bar q}} {\rmd^2\b} \right\rangle \left\langle\frac{\rmd \sigma_{q\bar q}} {\rmd^2\b'}\right \rangle=\langle \Sigma_{q\bar q}(\bmDelta)\rangle^2,
\ee
where
\be
\Sigma_{q\bar q}(\Delta)\equiv \int \rmd^2\b \,\rme^{-i\Delta\cdot \b}\,\frac{\rmd \sigma_{q\bar q}} {\rmd^2\b}
\ee
and the angular brakets denote average over the ground state wave function.

In the dense regime we have
\be\label{sigmaqqbexpo}
\frac{\rmd \sigma_{q\bar q}} {\rmd^2\b}=2\left( 1-\rme^{ -\frac{1}{2}\sigma_{\rm dip} \hat T_A(\b)}\right).
\ee
This is no longer a one-body operator, and its average can no longer be expressed in terms of the average of $\hat T_A(\b)$. However,  it is straightforward to sample the probability distribution associated with the ground state wave function, viz. $\vert\bra{\b_1,\cdots,\b_A}\Psi_0\rangle\vert^2$, and proceed to a numerical evaluation of the average of the cross section (\ref{sigmaqqbexpo}) (see next section). 

Similar considerations apply to the total diffractive cross section.   The elastic amplitude assumes that only the ground state of the nucleus contributes as an intermediate state. We can relax this assumption and allow  all possible diffraction eigenstates as intermediate states. Thus, in the weak field limit, we are led to define the total diffractive cross section as 
\be\label{averageSig2}
\int \rmd^2\b \,\rmd^2\b' \rme^{-i\bmDelta\cdot(\b-\b')} \, \sigma_{\rm dip}^2\sum_\alpha \left\vert  \bra{\alpha}\hat T_A(\b) \ket{\Psi_0}\right\vert^2 =\langle \Sigma^2_{q\bar q}(\bmDelta)\rangle,
\ee
where $\ket{\alpha}=\ket{\b_1,\cdots,\b_A}$ denotes a scattering eigenstate. By combining with Eq.~(\ref{coherentSigTA}), one finds that the incoherent cross section (in the weak field limit) is proportional to 
\be
\langle \Sigma^2_{q\bar q}(\bmDelta)\rangle-\langle \Sigma_{q\bar q}(\bmDelta)\rangle^2.
\ee
That is, the cross section is proportional to the fluctuation of the dipole cross section, a familiar result of the Good-Walker formalism \cite{GoodWalker1960,MiettinenPumplin1978}.

The evaluation of $\sum_\alpha \left\vert  \bra{\alpha}\hat T_A(\b) \ket{\Psi_0}\right\vert^2$ can conveniently be done in terms of the density-density correlation function. 
Considering $\vert\bra{\b_1,\cdots,\b_A}\Psi_0\rangle\vert^2$ as a symmetric probability distribution, we define the  associated one and two-point functions 
\be
\int\rmd^2\b_1\cdots\rmd^2 \b_A \vert \langle{\Psi_0}\ket{\b_1,\b_2,\cdots,\b_A}\vert^2 \,\sum_{i=1}^A \delta(\x - \b_i)=A \rho^{(1)}(\x), 
\ee
and 
\bea
&&\int\rmd^2\b_1\cdots\rmd^2 \b_A \vert \langle{\Psi_0}\ket{\b_1,\b_2,\cdots,\b_A}\vert^2 \,\sum_{i,j} \delta(\x - \b_i)\delta(\y - \b_j)\nonumber\\
&&\qquad\qquad=A \rho^{(1)}(\x) \delta(\x-\y)+A(A-1) \rho^{(2)}(\x,\y) .
\eea
These are normalized to unity: $\int\rmd^2 \x\,\rho^{(1)}(\x)=1$, $\int\rmd^2\x\, \rmd^2\y \,\rho^{(2)}(\x,\y)=1$. Note also that $\langle \hat\rho(\x)\rangle =A \rho^{(1)}(\x)$.
The density-density correlation function reads\footnote{We use a standard notation, $S(\x,\y)$, for the correlation function (see e.g. \cite{Blaizot_2014} for a discussion of this object in the context of Glauber Monte Carlo simulations). This should not be confused with the ${\cal S}$-matrix.}
\bea
&&S(\x,\y)=\langle \hat\rho(\x)\hat\rho(\y)\rangle-\langle \hat\rho(\x)\rangle\langle\hat\rho(\y)\rangle=A\rho^{(1)}(\x) \delta(\x-\y)\qquad\qquad\nonumber\\
 &&\qquad\qquad\qquad\qquad\qquad +A(A-1) \rho^{(2)}(\x,\y)-A \rho^{(1)}(\x)A \rho^{(1)}(\y).
\eea
In the absence of correlations between the particles, $\rho^{(2)}(\x,\y)=\rho^{(1)}(\x)\rho^{(1)}(\y)$ and 
\be
S(\x,y)=A\rho^{(1)}(\x) \delta(\x-\y)-A\rho^{(1)}(\x)\rho^{(1)}(\y).
\ee
The first term  represents the local fluctuation of the density. This is the source of the fluctuations responsible for the incoherent scattering. The last term in the formula above just ensures that $\int \rmd^2\x \, \,S(\x,\y)=0$, an equality which holds when the total number of nucleons does not fluctuate. 

We now return to the evaluation of the fluctuation of the nucleus profile $\hat T_A(\b)$, and note that 
\bea\label{fluctuTA}
&&\langle \hat T_A(\b) \hat T_A(\b')\rangle-\langle \hat T_A(\b)\rangle\langle \hat T_A(\b')\rangle=\int\rmd^2 \x \int\rmd^2 \y\, T_G(\x-\b)T_G(\y-\b' ) S(\x,\y)\nonumber\\
&&\qquad\qquad\simeq  A\int\rmd^2\x \rho^{(1)}(\x)T_G(\x-\b)T_G(\x-\b' )-\frac{1}{A} \langle \hat T_A(\b)\rangle\langle \hat T_A(\b')\rangle,
\eea
where, in the second line, we have assumed uncorrelated nucleons (hence the approximate equal sign).
 In Fourier space, Eq.~(\ref{fluctuTA}) yields 
 \be\label{incohdensdens}
 \int\rmd^2\b \, \rmd^2\b' \rme^{-i\bmDelta\cdot (\b-\b')} \left[  \int\rmd^2\x \,\rho(\x)T_G(\x-\b)T_G(\x-\b' ) -\frac{1}{A}\langle T_A(\b)\rangle\langle T_A(\b') \right].
\ee
 At small $\bmDelta$ the integration over the impact parameters becomes free, and the integral goes to zero. This implies that the fluctuations related to the random locations of the sources of the gluon field (here the nucleons) vanish.  This simple relation of the fluctuation of the cross section to the density-density correlation function holds strictly  in the dilute regime, where the cross section is given by Eq.~(\ref{dsigd2bTA}),  but should provide a reasonable approximation even in the dense regime\footnote{Note for instance that, for the relevant values of parameters, the expansion in $\sigma_{\rm dip}$ appears to converge rapidly, as indicated in Appendix \ref{sec:app1}.}. It explains the suppression of the geometrical  fluctuations observed in the incoherent cross section at small momentum transfer (see the discussion at the end of Sect.~\ref{sec:GFluct}). \\

The previous discussion concerned the scattering on a nucleus. One can extend it to the case of a proton, assimilating the valence quarks to the sources of the gluon field \cite{Mantysaari:2016ykx}. All we need to do is to replace the average density profile of the proton, well given phenomenologically by a Gaussian,  
\be
 T_G({\bf b}) = {1 \over 2 \pi B_G} e^{-{{\bf b}^2 / (2 B_G)}},
\label{eq:TpGaussian}
\ee
by the contribution of $N_q=3$ valence quarks
\be\label{TGc}
T_G({\bf b}) \mapsto \hat T_G(\b)={1 \over N_q} \sum_{i=1}^{N_q}T_q({\bf b}-{\bf b}_i).
\ee
Here
\be
T_q({\bf b}) = {1 \over 2 \pi B_q}e^{-{\bf b}^2/(2 B_q)}
\label{eq:Tq}
\ee
plays the role of $T_G$ earlier: it represents the average gluon density around a valence quark, with the parameter $B_q$ controlling the size of this gluon cloud.  In short, in this picture, valence quarks play the role of the nucleons, and the proton is considered as  a bound state of valence quarks.

\subsection{Fluctuations in the projectile}\label{sec:projfluct}

Aside from the geometrical fluctuations in the target that we have just discussed, one can also expect fluctuations in the 
projectile arising from the quantum mechanical splitting of the photon into a quark-antiquark pair: the resulting color dipole can have, event-be-event, a different size and orientation. 
The amplitude (\ref{eq:A_exclusive1}) is explicitly written as an average over these dipole degrees of freedom, the overlap of the photon and the vector meson wave functions playing the role of a distribution of dipole sizes \cite{MunierStastoMueller2001}. 
 However this overlap is not quite the square of a normalized wave function, and the ``non-diagonal'' character of this object  brings in subtleties that need to be dealt with.  We shall  keep the discussion general,  call $\psi_i$ the initial wave function (that of the photon) and $\psi_f$ the final one (that of the vector meson), and ignore the geometrical fluctuations of the proton.  By expanding $\psi_i$ and $\psi_f$ on the diffraction eigenstates (here color dipoles of various sizes), we get
\be
\ket{\psi_i}=\sum_\alpha C_i^\alpha \ket{\alpha}, \quad \ket{\psi_f}=\sum_\alpha C_f^\alpha \ket{\alpha},\quad \bra{\psi_f} \psi_i\rangle=\sum_\alpha C^{\alpha *}_{f} C_i^\alpha.
\ee
The elastic (or coherent) scattering amplitude is then given by 
\be\label{averageT2}
\bra{\psi_f} {\cal T}\ket{\psi_i}=\sum_\alpha C_{f}^{\alpha*} C_i^\alpha  \,t_\alpha,\qquad  {\cal T}\ket{\alpha}=t_\alpha \ket{\alpha}.
\ee
Here ${\cal T}$ stands for the (imaginary) scattering matrix and the states $\alpha$ are its eigenstates, the corresponding eigenvalues being denoted $t_\alpha$. 
To interpret  Eq.~(\ref{averageT2}) as an average, we divide $C^{f*}_\alpha C^i_\alpha$ by the overlap. We obtain 
\be
\label{averageT3}
\frac{\bra{\psi_f}{\cal T}\ket{\psi_i}}{\bra{\psi_f} \psi_i\rangle}= \sum_\alpha \frac{C_{f}^{\alpha *}C_i^\alpha }{\bra{\psi_f} \psi_i\rangle} t_\alpha= \sum_\alpha p_\alpha t_\alpha=\langle t\rangle,
\ee
where the probabilities $p_\alpha$ are properly normalised, $\sum_\alpha p_\alpha=1$. In terms of the ${\cal S}$-matrix, this relation takes the form 
\be\label{Sfi}
\bra{\psi_f}{\cal S}\ket{\psi_i}=\bra{\psi_f}\psi_i\rangle-\bra{\psi_f}{\cal T}\ket{\psi_i}= \bra{\psi_f}\psi_i\rangle\left[1-\frac{\bra{\psi_f}{\cal T}\ket{\psi_i}}{\bra{\psi_f}\psi_i\rangle}\right].
\ee
The overlap factor gives the contribution that exists in the absence of interaction. The second term within the squared brackets correctly measures the effect of the interactions  relative to the unit factor.   

Similarly, the total diffractive cross section is written in the following  way 
\be\label{inelastic2}
\frac{\sum_\alpha \bra{\psi_f}{\cal  T}\ket{\alpha}\bra{\alpha}{\cal T}^\dagger\ket{\psi_i}}{\bra{\psi_f} \psi_i\rangle}=  \sum_\alpha \frac{C_{f}^{\alpha *}C_i^\alpha }{\bra{\psi_f} \psi_i\rangle}t_\alpha^2= \langle t^2\rangle.
\ee
Leaving aside the Fourier transform, this expression is analogous to the sum over the diffraction eigenstates in  Eq.~(\ref{averageSig2}). However, aside from the normalization associated to the overlap, an important difference lies in the fact that in the target case, $\psi_i=\psi_f=\Psi_0$, with $\Psi_0$ the proton ground state. While $\psi_i$ and $\psi_f$ do refer to states with the same quantum numbers, Eq.~(\ref{inelastic2}) represents a non trivial extrapolation.

If one accepts Eq.~(\ref{inelastic2}), we are led to  expect  the incoherent diffractive cross-section to be   proportional to 
\be\label{incoherent6}
\frac{\bra{\psi_f}{\cal  T} {\cal  T}^\dagger\ket{\psi_i}}{\bra{\psi_f} \psi_i\rangle}-\left\vert\frac{\bra{\psi_f} {\cal  T}\ket{\psi_i}}{\bra{\psi_f} \psi_i\rangle}\right\vert^2=\langle t^2\rangle-\langle t\rangle^2,
\ee
where $\bra{\psi_f}{\cal  T} {\cal  T}^\dagger\ket{\psi_i}$ stands for $\sum_\alpha \bra{\psi_f}{\cal  T}\ket{\alpha} \bra{\psi_i}{\cal  T}\ket{\alpha}^*$.
This equation  has the correct property to vanish when $t_\alpha$ does not depend on $\alpha$. Indeed, when all the states $\alpha$ are absorbed at the same rate there can be no diffractive effects. 

Now,  in the calculation of the cross section, we need to reinstate the overlap in such a way that the coherent cross section is properly normalized. This implies  that one should multiply the result in Eq.~(\ref{incoherent6}) by the overlap squared.  Thus, to within overall numerical factors independent of $\psi_i$ and $\psi_f$, the incoherent diffractive cross section ends up being proportional to 
\be\label{incoherent2}
\bra{\psi_f}{\cal  T} {\cal  T}^\dagger\ket{\psi_i}\bra{\psi_f} \psi_i\rangle-\left\vert\bra{\psi_f} {\cal  T}\ket{\psi_i}\right\vert^2= \bra{\psi_f} \psi_i\rangle^2\left[\langle t^2\rangle-\langle t\rangle^2  \right].
\ee
Note that when $\psi_i=\psi_f=\Psi_0$ one recovers the standard result of the previous subsection.

\section{\label{sec:EbyEA} Event-by-event calculation}

We  turn now to the explicit calculation of the photoproduction cross section of vector mesons in the dipole model, Eq.~(\ref{eq:dxsection_coh1}), based on the explicit expression of the amplitude given in Eq.~(\ref{amplitudeA}). We specialize to the case of $J/\Psi$ production, and consider both coherent and incoherent contributions to the cross section. 

Let us start with a remark concerning the Fourier transform of the amplitude. In the previous section this was given by Eq.~(\ref{calADelta}). In fact there is a subtlety here, related to the definition of the impact parameter. It may be chosen as the transverse distance between the center of mass of the proton and that of the dipole, the latter coinciding with the location of the photon in the transverse plane. Now, the center of mass of the dipole depends on the fraction $z$ of the longitudinal momentum carried by one member of the dipole. If $\x_1$ and $\x_2$ denote respectively the transverse coordinates of the quark and that of the antiquark, as measured from the center of mass of the proton, then  $\r=\x_1-\x_2$ and $\b= z\x_1+(1-z) \x_2$. The ${\cal S}$-matrix element to be evaluated is therefore of the form ${\cal S}(\b+(1-z)\r,\b-z\r)$ which differs from the expression (\ref{Sbr})  in which  $\b$ is measured from the midpoint between the quark and the antiquark. However, we can write ${\cal S}(\b+(1-z)\r,\b-z\r))={\cal S}(\b'+\r/2,\b'-\r/2)$ with $\b'=\b+\left(\frac{1}{2}-z\right) \r$, and a simple change of variables allows us to write  the amplitude (\ref{calADelta}) in the form \footnote{The factor $1/2-z$ in Eq.~(\ref{eq:sqqintegral}) differs from the factor $1-z$ often used in that context (see e.g. \cite{IPSat2006,Mantysaari:2016ykx}, including our previous work \cite{TrainiBlaizot2019}). The present discussion is inspired from Ref.~\cite{Hatta_2017}. }
\be
\Sigma_{q\bar q}({\xP,\br,z, \bmDelta })  = \int d^2 {\bf b}{{d \sigma_{q \bar q} \over d^2 {\bf b}}}({\bf b},{r}, {\xP})\, e^{-i({\bf b}-[(1/2-z) {\bf r}]) \cdot {\bmDelta}},
\label{eq:sqqintegral}
\ee
where the variable conjugate to $\bmDelta$ is not simply $\b$ as  in (\ref{calADelta}), but $\b-(\frac{1}{2}-z)\r$. Note that $\Sigma_{q\bar q}({\xP,\br,z, \bmDelta })$ differs from the amplitude ${\cal A}_{q\bar q}(\xP,\r,\bmDelta)$ solely by the $z$-dependent phase factor. 

The incoherent photo-production cross section can be written in the form 
\bea\label{crosssection}
 {16 \pi \over (1+\beta^2)}\,{d \sigma^{\gamma p \to V p'} \over dt} = \bigg<\left\vert  {\cal A}^{\gamma p \to V p'}({\xP},\bmDelta)\right\vert ^2 \bigg> - 
\left\vert  \bigg< {\cal A}^{\gamma p \to V p'}({\xP}, \bmDelta)\bigg>\right\vert ^2\nonumber\\
\eea
where the second term give the elastic, or coherent, cross section (see Eq.~(\ref{eq:dxsection_coh1})). The angular brackets represent the various averages over the projectile and the target. These averages will be calculated by sampling appropriate distributions, as will be explained in the following subsections.

\subsection{\label{sec:GFluct}Proton shape fluctuations}

In this subsection, we review  the effects of proton shape  fluctuations, ignoring the fluctuations of the dipole sizes. Following \cite{M&Schenke2016} we assume that gluons are distributed around the valence quarks, geometrical fluctuations arising from the random locations of the valence quarks, event-by-event. To deal with these, we replace the average density profile of the proton, $T_G(b)$, by the expression  (\ref{TGc}) involving the sum of gluon densities attached to the valence quarks.\footnote{Recently~\cite{TrainiBlaizot2019} an approach connecting the density profile to the valence quark distribution in the proton has been proposed. The approach does not need any free parameters even for the gluon distribution which is directly connected to the valence quark density by means of  appropriate DGLAP evolution.}
Typical parameters for the various Gaussian distributions involved in Eqs.~(\ref{eq:TpGaussian}, \ref{TGc}, \ref{eq:Tq}) are respectively $B_G=4 \;{\rm GeV}^{-2}$, $B_{qc}=3.3\; {\rm GeV}^{-2}$, and $B_q=0.7\;{\rm GeV}^{-2}$ (note that $B_G=B_{qc}+B_q$). In terms of root mean square radii $\sqrt{2B}$, these numbers correspond to $0.56\; {\rm fm}, 0.51 \;{\rm fm}$ and $0.24\; {\rm fm}$ respectively for the radius of the proton, that of the valence quark distribution and the size of the gluon cloud around each valence quark. A sampling procedure selects randomly a large number of quark positions ${\bf b}^{\{c\}}_i$ in a Gaussian distribution with parameter $B_{qc}$. A given choice of ${\bf b}^{\{c\}}_i$ is referred to as a configuration. The cross sections are then obtained by adding the contributions of each configuration, and dividing by the total number $N_C$ of these configurations (see Eq.~(\ref{crosssectionb}) below). The results presented below were obtained for $N_C=10 \,000$  configurations,  as in ref.~\cite{TrainiBlaizot2019}.

The dipole cross section depends on the configuration, and we write
\bea
&& {d \sigma_{q \bar q}^{\{c\}} \over d^2 {\bf b}} (\xP,{\bf b},{\bf r}) = 
2 \left[1-\exp\bigg(- {1\over 2 } \sigma_{\rm dip}(\xP,\r)\,T^{\{c\}}_G({\bf b}) \bigg) \right], 
\label{eq:dXsqbarqGF}
\eea
where $T^{\{c\}}_G({\bf b})$ is evaluated according to Eq.~(\ref{TGc}) in which $\b_i\mapsto \b_i^{\{c\}}$. The Fourier transform of this expression, according to Eq.~(\ref{eq:sqqintegral}), yields $\Sigma^{\{c\}}_{q\bar q}({\xP,\br,z, \bmDelta })$. 
 The integration over the impact parameter involved in the Fourier transform is done as explained in Appendix~\ref{sec:app1}. 
The scale $\mu$ in the gluon distribution function ${\xP} g({\xP},\mu^2)$ that enters $\sigma_{\rm dip}(\xP,\r)$ (see Eq.~(\ref{sigmadip})) is related to the size ${ r}$ of the dipole \cite{IPSat2002}
\be
\mu^2 = \mu^2({r}^2) = \mu_0^2 + {4 \over {r}^2},
\label{eq:mu} 
\ee
and the gluon distribution is parameterized as 
\be
\xP g(\xP,\mu_0^2) = A_g \, \xP^{-\lambda_g}\,(1-\xP)^{5.6}. 
\label{eq:xg}
\ee
We use the parameters given in Ref.~\cite{IPSat2002}.

Since, in this subsection, we are ignoring the fluctuations in the dipole degrees of freedom, we define an average $\bar\Sigma$ by integrating over $\r$ and $z$, 
\be 
\bar \Sigma^{\{c\}}_{q\bar q}(\bmDelta )=\int d^2 {\bf r}   \int {dz \over 4 \pi} \left[\Psi^*_\gamma \Psi_V \right]_{({\bf r},z)}\Sigma^{\{c\}}_{q\bar q}({\br,z, \bmDelta }). 
\ee
Physically $\bar\Sigma$ represents the cross section of an average dipole scattering on a fluctuating proton. 
In terms of $\bar \Sigma$, 
the incoherent cross section (\ref{crosssection}) takes then the simple form
\bea\label{crosssectionb}
 {16 \pi \over (1+\beta^2)}\,{d \sigma^{\gamma p \to V p'} \over dt} ={1 \over N_C} \sum_{\{c\}} 
\bigg\vert    \bar
\Sigma^{\{c\}}_{q\bar q}(\br,z, \bmDelta )\bigg\vert^2- \left\vert {1 \over N_C} \sum_{\{c\}} \bar \Sigma^{\{c\}}_{q\bar q}({\br,z, \bmDelta })\right\vert^2 ,\nonumber\\
\eea
where the sums run over the configurations defined above. The coherent cross section is given by the second term in this equation.

\begin{figure}[h!]
\vspace{-10.0em}
\centering\includegraphics[width=0.85\columnwidth,clip=true,angle=0]{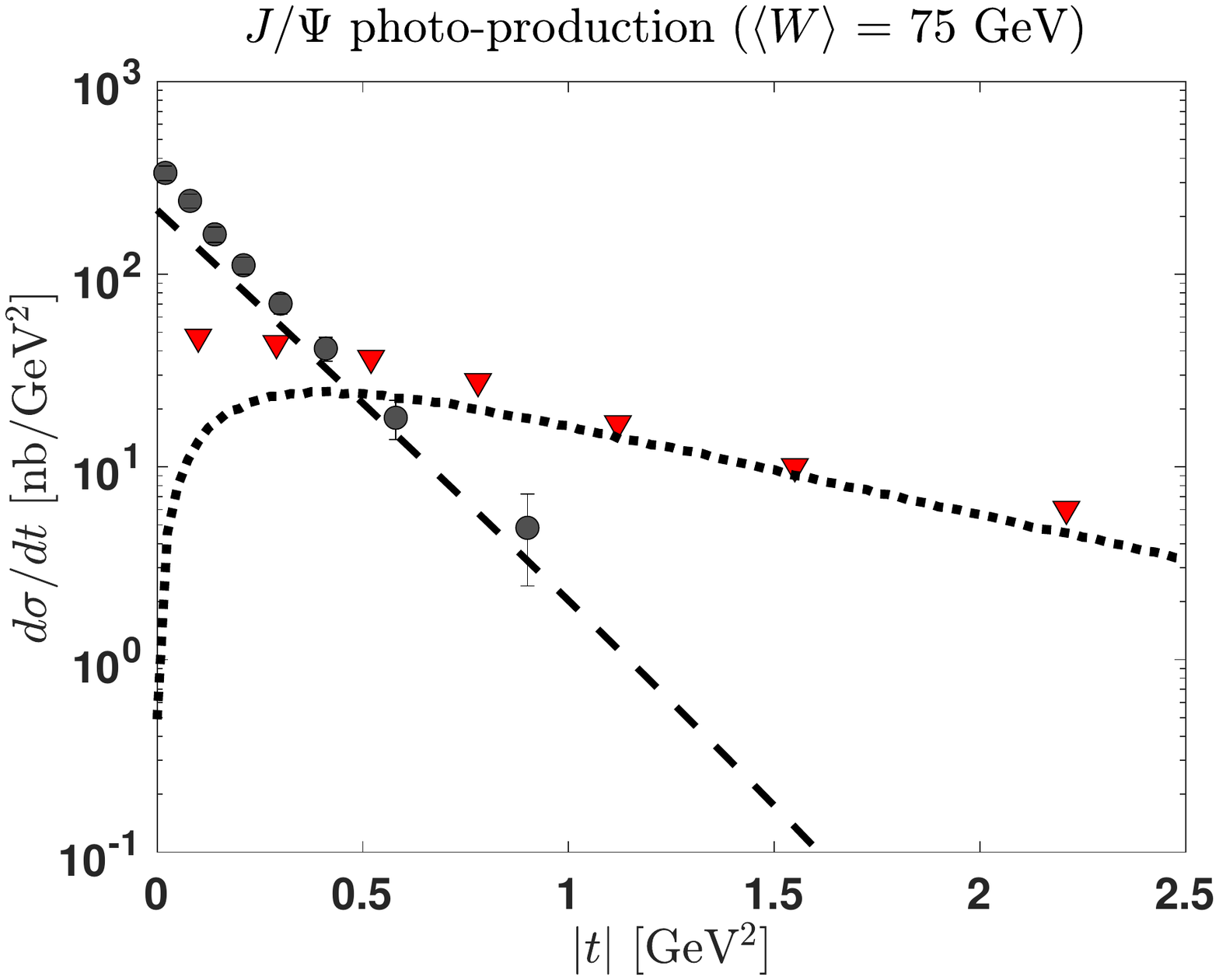}
\vspace{-7.0em}
\caption{\small  (color on line) Coherent and incoherent cross sections for $J/\Psi$ photoproduction at $W = 75$ GeV.  The dashed and dotted lines represent respectively the coherent and incoherent cross sections calculated according to Eq.~(\ref{crosssectionb}). Coherent H1 data from Refs. \cite{H1JPsi1,H1JPsi2} (circles), incoherent data (triangles) from H1 and ZEUS experiments of Refs. \cite{H1JPsi2,ZEUS1}.}
\label{fig:GF_sampling_JPsi_Q20p0_TG_W75_incoh_coh}
\end{figure}

Fig.~\ref{fig:GF_sampling_JPsi_Q20p0_TG_W75_incoh_coh} shows the results of calculations of both coherent and incoherent cross sections for HERA kinematics and $J/\Psi$ photo-production. Only the fluctuations of the proton shape are taken into account in the calculation. 

At small momentum transfer $\bmDelta$, the cross section is dominated by the elastic component. This decreases exponentially, $\propto \rme^{-\vert t\vert B_D}$ where the parameter $B_D$ is of the order of $B_G$ (but slightly larger \cite{Caldwell:2010zza}). That is, the exponential drop of the coherent cross section is essentially determined by the size of the gluon cloud in the proton. This is clear from Eq.~(\ref{dsigd2bTA}) showing that in the weak field limit, the average $\langle\Sigma(\bmDelta)\rangle$ is proportional to the Fourier transform of the average $\langle \hat T_G(\b)\rangle$.   

The coherent contribution drops rapidly, and ceases to be dominant when the geometrical fluctuations start to become important, which occurs when $|t|\gtrsim 0.4-0.5$ Gev$^2$, corresponding to a typical length scale of order 0.3 fm. The incoherent cross section decreases with a much smaller slope than the coherent cross section, the latter becoming rapidly negligible as $\Delta$ increases. The good agreement with the data confirms that
 geometric fluctuations associated to a  ``lumpy''  proton appears to be a crucial ingredient to reproduce the incoherent cross section~\cite{M&Schenke2016} at moderate momentum transfer, $\vert t\vert\lesssim 2.5$ GeV$^2$.

The same Fig.~\ref{fig:GF_sampling_JPsi_Q20p0_TG_W75_incoh_coh} reveals that, at very low momentum transfer, the effect of geometrical fluctuations tends to vanish. This feature, already visible in the early model of  Ref.~\cite{MiettinenPumplin1978}, is in fact common to most hot spot models. It has a simple origin.  The gluon density ``seen'' by a dipole at a given impact parameter $\bb$, fluctuates event-by-event depending on whether there are other valence quarks in the vicinity of $\bb$. However, at small momentum transfer, $\bb$ is averaged over areas of the order or larger than the proton size. As a result, the dipole ``sees'' an average proton density profile, in which density fluctuations are averaged out. A more formal argument relies on the properties of the density-density correlation function that we have recalled in Sect.~\ref{flucttarget}  (see Eq.~(\ref{incohdensdens})). Note that if one allows the total number of nucleons $A$ to fluctuate, one gets a contribution at $\bmDelta=0$. Indeed, in this case, 
\be
\int\rmd^2\b \, \rmd^2\b' \,\left[\langle \hat T_A(\b) \hat T_A(\b')\rangle-\langle \hat T_A(\b)\rangle\langle \hat T_A(\b')\rangle\right]=\langle A^2\rangle-\langle A\rangle^2.
\ee
This observation may be exploited in hot spot models to compensate for the vanishing of the incoherent cross section at vanishing $\Delta$.   
However, this is not needed since, as we show in the next section,  precisely that region of small momentum transfer is dominated by fluctuations of the dipole size in the initial state.

\subsection{\label{sec:rFluct}Dipole size fluctuations}

\begin{figure}[h!]
\vspace{-0.0em}
\centering\includegraphics[width=0.75\columnwidth,clip=true,angle=0]{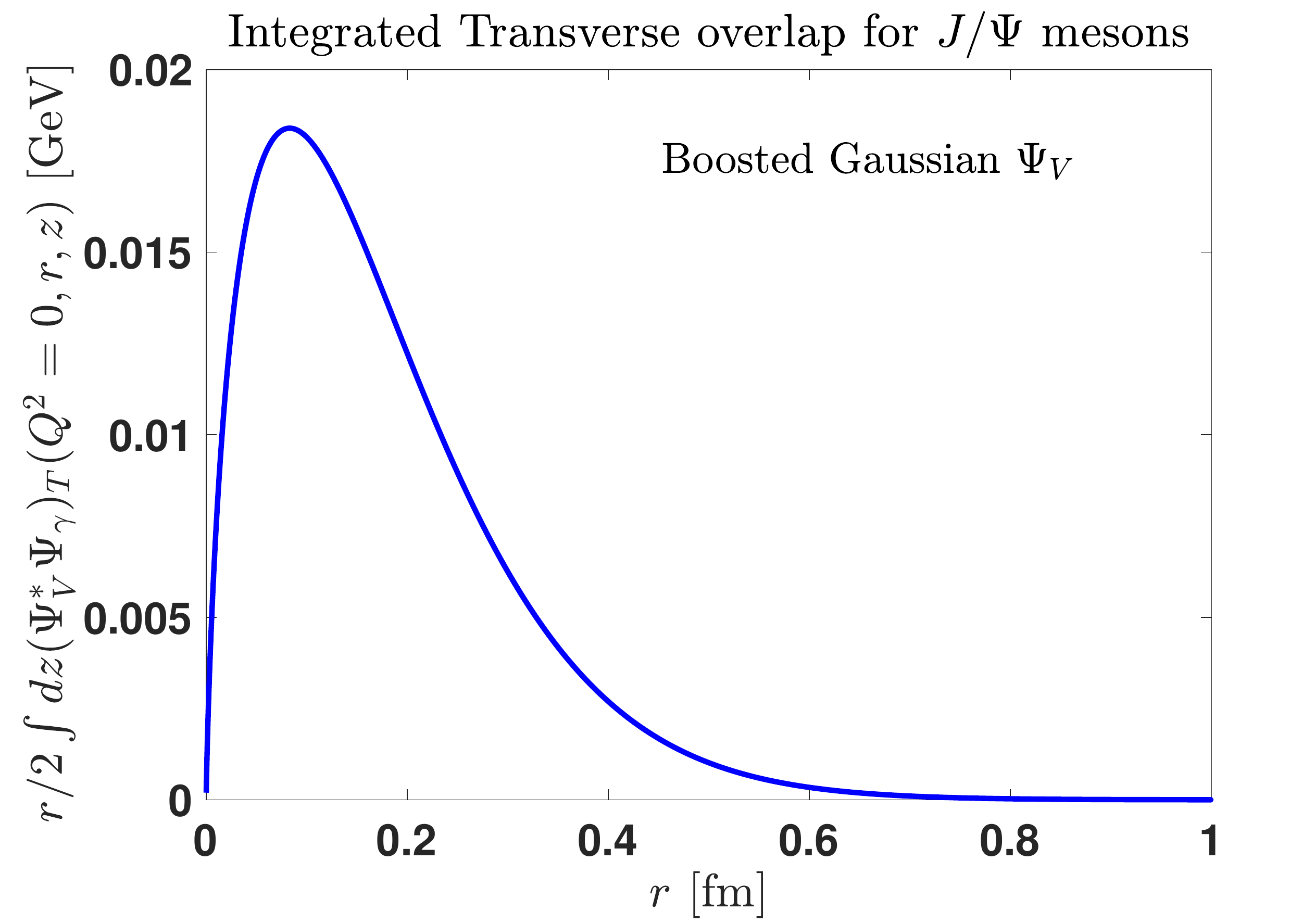}
\vspace{-0.0em}
\caption{Photon - $J/\Psi$ transverse overlap function~(\ref{eq:intz2b}) integrated over $z$.}
\label{fig:OTJ_Psi}
\vspace{+0.0em}
\end{figure}

In order to calculate the fluctuation of the dipole size, we assimilate the overlap between the photon and the $J/\Psi$ meson to a probability distribution, as we have already mentioned. More precisely, we set 
\be
{\cal P}(r) \propto  {2 \pi r} \int {dz \over 4 \pi} \, [\Psi^*_V\Psi_\gamma]_{(r ,z)}, \qquad \int dr {\cal P}(r) = 1.
\label{eq:intz2b}
\ee
That is, we define the probability ${\cal P}(r)$ after integration over $z$. Since we shall keep the $z$-average intact, this introduces a slight inconsistency in the calculation which is of minor relevance since the effect of the $z$-dependence is in any case small. The distribution ${\cal P}(r)$ is plotted in Fig.~\ref{fig:OTJ_Psi}. This distribution is  sampled in order to get an event-by-event realization of   random  dipole sizes.  In our calculation, we use  $N_C=10\, 000$ configurations chosen with the sampling procedure detailed in Appendix~\ref{sec:samplingrFluct1}.

At this point, in order to make the calculation clearer, we ignore temporarily the proton shape fluctuations, that is, $\Sigma$ in Eq.~(\ref{eq:dxsection_incoh_fluct2}) below is averaged over the proton ground state. In fact we shall use the simple Gaussian approximation in which $\Sigma$ is calculated with the average profile $T_G(b)$. We shall denote $\tilde \Sigma$ the corresponding average of $\Sigma$. That is $\tilde \Sigma$ is calculated by substituting in Eq.~(\ref{eq:sqqintegral}) the expression (\ref{sigmadipexp}) for $\rmd \sigma/\rmd^2\b$.

Each event (which we shall also refer to as a configuration)  corresponds to a choice of a dipole size  $r_\cc = \vert {\bf r}_\cc\vert $  extracted from the probability density (\ref{eq:intz2b}) with the appropriate frequency (see Fig.~\ref{fig:OTJ_Psi}). The  average of the amplitude over the dipole degrees of freedom ($\r$ and $z$) can therefore be written as 
\bea
\bigg< {\cal A}^{\gamma p \to V p'}( \bmDelta)\bigg>&=& \int d^2 {\bf r}   \int {dz \over 4 \pi} \left[\Psi^*_\gamma \Psi_V \right]_{({\bf r},z)}\tilde\Sigma_{q\bar q}({\br,z, \bmDelta })\nonumber\\ 
&\simeq& {1 \over N_C} \sum_{\{c\}}  \int d^2 {\bf r}   \int {dz \over 4 \pi} \left[\Psi^*_\gamma \Psi_V \right]_{({\bf r},z)}\tilde \Sigma_{q\bar q}({\br_\cc,z, \bmDelta }).\nonumber\\
 \label{eq:dxsection_incoh_fluct2}
\eea
The second line makes explicit that neither the average over $z$ nor the angular average over the orientation of the dipole are part of the ``configurations'' which, in the present case just represent the dipole sizes. The average over the dipole size is performed via the sum over the configurations, that is over a selected sample of dipole sizes. The approximate equal sign in the second line emphasizes that this calculation is not quite exact since, as indicated earlier, the sampling is done with a probality distribution that is integrated over $z$ rather than with a $z$ dependent distribution. The $r$-integration in the second line of the equation above does not affect $\Sigma_{q\bar q}({\br_\cc,z, \bmDelta }) $. Although this may seem confusing at first sight, it is convenient to keep the writing as it is because it exhibits the proper normalization of the amplitude, as explained in Sect.~\ref{sec:projfluct}. 

A similar procedure is followed for the total diffractive cross section, which we may write as follows
\bea
\bigg<\left\vert  {\cal A}^{\gamma p \to V p'}(\bmDelta)\right\vert ^2 \bigg> \simeq {1 \over N_C} \sum_{\{c\}} 
\bigg\vert    \int d^2 {\bf r}   \int {dz \over 4 \pi}
 \left[  \Psi^*_\gamma \Psi_V \right]_{({\bf r},z)} 
\tilde\Sigma_{q\bar q}(\br_\cc,z, \bmDelta )\bigg\vert^2.\nonumber \\
\label{eq:dxsection_incoh_fluct3}
\eea
As before the role of the $r$-integration is to ensure the proper normalization. It does not affect $\tilde\Sigma(\br_\cc)$, which, we recall, is averaged over the ground state density of the proton. 

\begin{figure}[h!]
\vspace{-10.0em}
\centering\includegraphics[width=0.75\columnwidth,clip=true,angle=0]{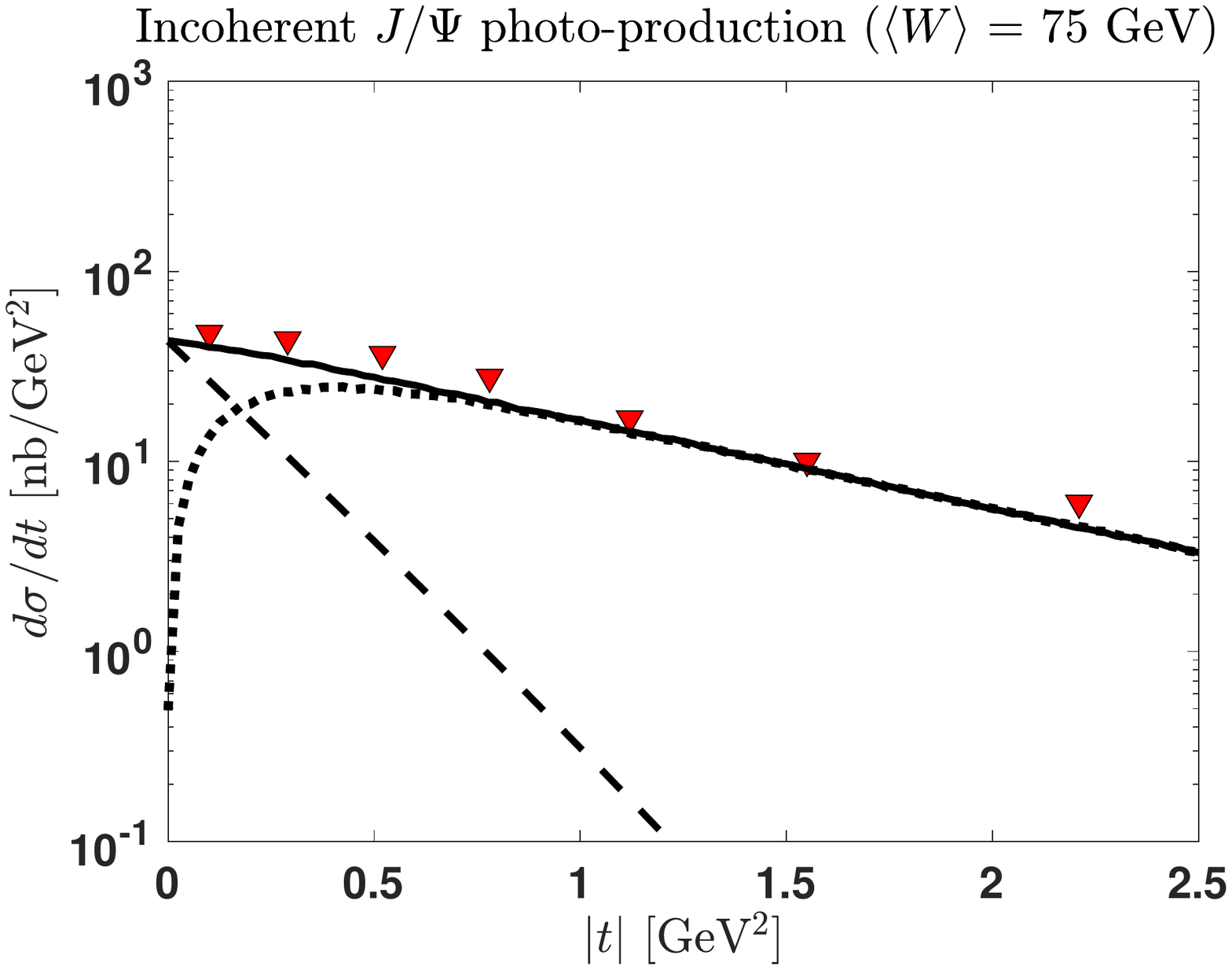}

\vspace{-7.0em}
\caption{\small  (color on line) Incoherent $J/\Psi$ photo-production cross section at the HERA (H1) kinematics $W = 75$ GeV. The dashed and dotted lines represent the incoherent cross sections calculated respectively from Eqs.~(\ref{eq:dxsection_incoh_fluct2}) and (\ref{eq:dxsection_incoh_fluct3}) for the dipole size fluctuations,  and Eq.~(\ref{crosssectionb}) for the proton shape fluctuations.  The full line represents the total incoherent cross section, estimated here as the sum of the two (we assume that the two types of fluctuations are only weakly correlated). 
Data as in Fig.~\ref{fig:GF_sampling_JPsi_Q20p0_TG_W75_incoh_coh}.}
\label{fig:rOF_sampling_GF_JPsi_Q20p0_TG_W75_incoh}
\end{figure}

The contribution of dipole size fluctuations to the incoherent diffractive cross section in then obtained from the general expression (\ref{crosssection}), in which the needed averages are given in Eqs.~(\ref{eq:dxsection_incoh_fluct2}) and (\ref{eq:dxsection_incoh_fluct3}). 
An explicit numerical calculation  based on these equations is shown in 
Fig.~\ref{fig:rOF_sampling_GF_JPsi_Q20p0_TG_W75_incoh}. The result is compared to the contribution of the geometrical shape fluctuations as estimated in the previous subsection, as well as to the data on  $J/\Psi$ photo-production obtained at HERA. The effects of size fluctuations  are substantial at low momentum transfer while for $\vert t\vert \gtrsim 0.5$ Gev$^2$ their contribution is hidden under that of  the gluon shape fluctuations. As the plot reveals, the inclusion of the   dipole size fluctuations allows us to reproduce the data in the region of small momentum transfer. Given that the calculation contains no additional parameter, this agreement is  gratifying.

Dipole size fluctuations induce fluctuations in the dipole cross section. This makes the present discussion reminiscent of the fluctuating cross sections introduced in \cite{Blaettel:1993rd,Blaettel:1993ah}. It is also clear, from Eq.~(\ref{SQs}) for instance, that fluctuations in the dipole size can be viewed as fluctuations of the saturation momentum of the proton. In fact, 
in Ref.~\cite{M&Schenke2016} $Q_s$-fluctuations have been considered within the IP-Glasma framework, guided by the interpretation of multiplicity distributions and rapidity correlations in $p+p$ collisions in terms of saturation scale fluctuations~\cite{McLerranTribedy2016,BzdakDusling2016}. After some adjustment of the $Q_s$-distribution, a reasonable agreement with the data was achieved \cite{M&Schenke2016}. As mentioned above, varying the number of hot spots in hot spot models can also generate a contribution at small momentum transfer. We believe however that the  dipole size fluctuations discussed in the present paper provide a more natural explanation, and can be estimated without any additional parameters or adjustments. Together with the transverse proton shape fluctuations, it provides a  consistent picture of the incoherent cross section for moderate momentum transfers.

\section{\label{sec:conclusions}Summary and conclusions}

In this  paper we have considered the photo-production of $J/\Psi$  mesons within the dipole model. We have extended the approach of Ref.~\cite{M&Schenke2016,TrainiBlaizot2019} dealing with event-by-event fluctuations of the proton shape to include the fluctuations of the $q \bar q$ dipole size. We have shown that the latter are significant at small momentum transfer, where the impact parameter of the dipole is integrated almost freely over the size of the proton. The dipole  then sees essentially an average proton, in which the shape fluctuations of the proton are averaged out (as follows from a simple argument based on the analysis of the density-density correlation function). We have suggested a way to treat the fluctuations of the size of the  dipole  in which the splitting of the incoming virtual photon into a quark - antiquark pair is a quantum mechanical event to  which we can associate a probability, proportional to the overlap between the incoming photon and the produced vector meson wave functions. We have shown that this simple procedure allowed us to get an excellent agreement with HERA data, without adjustment of any additional parameters. We have noted that dipole size fluctuations could be interpreted as fluctuations of the saturation momentum scale of the proton. However, the phenomenological study presented in this paper offers a simple and unified picture of the role of geometrical fluctuations associated to  natural degrees of freedom, namely the valence quarks of the proton and the (heavy) quark components of the light cone wave functions of the photon and the $J/\Psi$ meson.

\begin{acknowledgements}

M.T. thanks the members of the Institut de Physique Th\'eorique, Universit\'e Paris-Saclay (CEA), for their warm hospitality during a visiting period when the present study was initiated. He thanks also the Physics Department of Valencia University for support and friendly hospitality. J.P.B. thanks the ECT* in Trento for hospitality during a visit that allowed this project to be completed. 

\end{acknowledgements}

\appendix

\section{\label{sec:app1} Integration over impact parameter}

We consider here the Gaussian approximation, with  the proton  profile given by Eq.~(\ref{eq:TpGaussian}),  $T_G({\bf b}) = 1/(2 \pi B_G) e^{-{\bf b}^2/(2 B_G)}$. The generalization to other cases is straightforward. The Fourier transform  (\ref{eq:sqqintegral}) of the dipole cross section reads (ignoring here the $z$ dependence which factorizes) 
\bea
\Sigma(\xP, \r, \bmDelta)  &=&  \int d^2 {\bf b}\,\rme^{-i{\bf b} \cdot {\bmDelta}} \, {{d \sigma_{q \bar q} \over d^2 {\bf b}}}({\bf b},{\bf r}, x_{\xP})  \nonumber \\
&=&2 \int d^2 {\bf b} \,\rme^{-i{\bf b} \cdot {\bmDelta}} \,\,  \bigg[1\!-\!\exp\left(\! - {1 \over 2 } \sigma_{\rm dip}(\xP,\r)\,T_G({\bf b}) \!\right)\!\bigg] \nonumber \\
&=&2\int d^2 {\bf b} \,\rme^{-i{\bf b} \cdot {\bmDelta}} \,\,  \bigg[1\!-\!\exp\left(\!- a(\xP,r) \, {1 \over 2 \pi B_G} e^{-{{\bf b}^2 / (2 B_G)}\!}\right)\!\bigg] \nonumber \\
\eea
with $a(\xP,r) = {\pi^2 \over 2 N_c} {\bf r}^2 \alpha_S(\mu^2(r))\,{\xP} g({\xP},\mu^2(r))=\sigma_{\rm dip}/2$. After expanding $\big[1-\exp(...)\big]$ in powers of $a$,
one performs the integration over the impact parameter and obtains
\bea
 \Sigma (\xP, r, \bmDelta^2) &=& -2 \sum_{n=1}^\infty \,{1 \over n!}\,{[-a(\xP, r)]^n \over [2 \pi B_G]^n}\,\int d^2 {\bf b} \,\, e^{-{{n \over 2 B_G}{\bf b}^2}}\,e^{-i{\bf b} \cdot {\bmDelta}} \nonumber \\
&=& - 2 \cdot (2 \pi B_G) \sum_{n=1}^\infty {1 \over n}{1 \over n!}\,{[-a(\xP,r)]^n \over [2 \pi B_G]^n}\,e^{-{{B_G \over 2 n}{\bmDelta}^2}} \nonumber \\
& \approx& - 2 \cdot (2 \pi B_G) \sum_{n=1}^{n_{max}} {1 \over n}{1 \over n!}\,{[-a(\xP,r)]^n \over [2 \pi B_G]^n}\,e^{-{{B_G \over 2 n}{\bmDelta}^2}}.
\label{eq:Iexp}
\eea
In the last line,  the sum has been truncated at a maximum value ${n_{max}}$  
(for the $J/\Psi$ meson the choice  $n_{max} = 4$ ensures an excellent convergence).

\section{ Overlap functions $[{\Psi^*_V\Psi_\gamma}]_{T,L}$}\label{sec:overlapPSI}

In this paper we use the same wave functions as in our previous work \cite{TrainiBlaizot2019}. We just recall here the form of the relevant overlap for the photo-production of $J/\Psi$ mesons (see also \cite{IPSat2006,FSS2004} for more details). 
We have 
\bea
&& [\Psi^*_V\Psi_\gamma]_{({\bf r},z)} = {e \over \pi \sqrt{2}} {N_c \over z (1-z)}\left\{ m^2_f K_0(\epsilon r) \phi_T(r,z) \right.\nonumber \\
&& \qquad\qquad\qquad - \left. \left[z^2 + (1-z)^2 \right]\epsilon K_1(\epsilon r) \partial_r \phi_T\right\} 
\label{eq:overlapT}
\eea
where $r=\vert {\bf r}\vert $, $e=\sqrt{4 \pi \alpha_{\rm em}} \approx \sqrt{4 \pi /137}$,   $\epsilon=m_f$ the charm quark mass, and $N_c = 3$ is the number of colors. Finally $K_0$ is the modified Bessel function of second kind, and $\partial_r K_0(\epsilon r) = - \epsilon K_1(\epsilon r)$. 
The boosted Gaussian wave function in configuration space is written as follows \cite{FSS2004,NNPZ94_97_1,NNPZ94_97_2}
\bea
\phi_{T}(r,z) ={\cal N}_{T}\,z(1-z) \exp \left[- {m^2_f {\cal R}^2 \over 8 z (1-z)} 
 + {m^2_f {\cal R}^2 \over 2} \right]  \exp \left[ - {2 z (1-z) r^2 \over {\cal R}^2} \right] \nonumber \\
\label{eq:boostedG}
\eea
and ${\cal N}_{T}$ and ${\cal R}$ are fixed by normalization conditions and the decay width \cite{IPSat2006}. The values of the relevant parameters are ${\cal N}_{T}=0.578$, ${\cal R}^2=2.3$ Gev$^{-2}$ and $m_f=1.4$ Gev.

\section{Sampling procedure for $r$-Fluctuations}\label{sec:samplingrFluct1}

In order to sample the transverse overlap function displayed in Fig.~\ref{fig:OTJ_Psi}, we first express it in a form which is mathematically convenient for the statistical sampling. This is achieved by a fit of the overlap with the sum of two Gaussian functions:
\bea
 {2 \pi r} \int {dz \over 4 \pi} \,[\Psi^*_V\Psi_\gamma]_{(r, z)}  &=& a_1 e^{-(r-b_1)^2/c_1^2} + a_2 e^{-(r-b_2)^2/c_2^2} \nonumber \\
&=& f_{1G}(r) + f_{2G}(r) \approx  F_{G}^{(2)}(r),\label{eq:fitG2}
\eea
with parameters given in table~\ref{tab:DUE}.
\begin{table}[h!]
\caption{Numerical values of the parameters in  Eq.~(\ref{eq:fitG2})}
\begin{center}
\begin{tabular}{c c c c c c}
\hline
\hline
\\
$a_1$ & $b_1$ & $c_1$ & $a_2$ & $b_2$ & $c_2$ \\
\\
0.01114 & 0.07989 & 0.08078 & 0.0105 & 0.1825 & 0.1784 \\
\\
\hline
\hline
\end{tabular}
\end{center}
\label{tab:DUE}
\end{table}
After normalisation, one obtains a probability distribution 
\bea
{\cal P}_{G}^{(2)}(r)  & = & {1 \over \int dr F_{G}^{(2)}(r)} \, F_{G}^{(2)}(r) 
\label{eq:fit_NORM}\\
& = & a {1 \over \sqrt{\pi c_1^2}}e^{-{(r - b_{1})^2 \over c_1^2}} + (1-a) {1 \over \sqrt{\pi c_2^2}}e^{-(r - b_{2})^2 \over c_2^2};\nonumber \\
\label{eq:rxry_fit}
\eea
where $a\approx0.32$.

The sum of the two Gaussian functions in Eq.~(\ref{eq:fit_NORM}) can then be sampled by a random selection of a single term of the sum, followed by the sampling of the distribution associated to that term~\cite{Kaloseta2008}.

\section{\label{sec:app2}Phenomenological corrections}

The derivation of the amplitude for the exclusive vector meson production 
(or DVCS amplitude if $V \to \gamma$, real photon) relies on the assumption that the amplitude ${\cal A}$ of  Eq.(\ref{eq:A_exclusive0}) purely imaginary. The corrections due to the presence of the real part is accounted for  by the factor $(1+\beta^2)$ multiplying the differential cross sections (\ref{eq:dxsection_coh1}). $\beta$ is the ratio of real and imaginary parts of the amplitude and is calculated  as follows \cite{IPSat2006}
\be
\beta = \tan {\pi \lambda \over 2}; \;\;\;\;\; {\rm with}\;\;\;\; \lambda \equiv {\partial \ln ({\rm Im }\,{\cal A}_{T,L}^{\gamma^* p \to V p}) \over \partial \ln(1/x_{\xP})}.
\ee
In addition for vector meson production (or DVCS) one should use the off-diagonal (or generalized) gluon distributions \cite{off-diagG}. Such a``skewness'' effect is accounted for (in the limit of small $x_{\xP}$), by multiplying the gluon distribution $\xP g(\xP,\mu^2)$ by a factor $R_g$ given by \cite{IPSat2006}
\bea
R_g(\lambda_g) & = & {2^{\lambda_g+3} \over \sqrt \pi}{\Gamma(\lambda_g+5/2) \over \Gamma(\lambda +4)},
\nonumber \\
{\rm with}\;\;\;\;  \lambda_g & \equiv & {\partial \ln [x_{\xP} g(x_{\xP},\mu^2)]\over \partial \ln (1/x_{\xP})}.
\eea

%

%

\end{document}